\documentclass[aps,prl,twocolumn,superscriptaddress,notitlepage]{revtex4-1}
\usepackage[utf8]{inputenc}
\usepackage[
colorlinks=true,
urlcolor=blue,
linkcolor=blue,
citecolor=magenta
]{hyperref}
\usepackage{amsmath,amsfonts,amssymb}
\usepackage{graphicx}
\usepackage{xcolor}
\usepackage{esint}

\begin{document}

\title{Low frequency Raman response near Ising-nematic quantum critical point: a memory matrix approach}
\author{Xiaoyu Wang}
\affiliation{National High Magnetic Field Laboratory, Tallahassee, FL 32310, USA}
\author{Erez Berg}
\affiliation{Department of Condensed Matter Physics, Weizmann Institute of Science, Rehovot 76100, Israel}
\begin{abstract}
 Recent Raman scattering experiments have revealed a ``quasi-elastic peak" in $\mathrm{FeSe_{1-x}S_x}$ near an Ising-nematic quantum critical point (QCP)~\cite{zhang17}. Notably, the peak occurs at sub-temperature frequencies, and softens as $T^{\alpha}$ when temperature is decreased toward the QCP, with $\alpha>1$. In this work, we present a theoretical analysis of the low-frequency Raman response using a memory matrix approach. We show that such a quasi-elastic peak is associated with the relaxation of an Ising-nematic deformation of the Fermi surface. Specifically, we find that the peak frequency is proportional to $ \tau^{-1}\chi^{-1}$, where $\chi$ is the Ising-nematic thermodynamic susceptibility, and $\tau^{-1}$ is the decay rate of the nematic deformation due to an interplay between impurity scattering and electron-electron scattering mediated by critical Ising-nematic fluctuations. We argue that the critical fluctuations play a crucial role in determining the observed temperature dependence of the frequency of the quasi-elastic peak. 
 At frequencies larger than the temperature, we find that the Raman response is proportional to $\omega^{1/3}$, consistently with earlier predictions~\cite{klein18a}. 
\end{abstract}
\maketitle
Many unconventional superconductors, such as the iron-based superconductors and hole-doped cuprates, host an Ising-nematic phase where the discrete crystalline rotational symmetry ($C_{4}$) is spontaneously broken \cite{Daou_2010,Chu:2010aa,Kohsaka:2007aa,bohmer16,coldea18,reiss18}. Upon doping or pressure, the nematic transition temperature is suppressed to zero, pointing to a putative Ising-nematic quantum critical point (QCP), where the nematic susceptibility diverges \cite{hosoi2016,Urata:2016aa,reiss17}. Close to the QCP, novel non-Fermi liquid behaviors have been observed, as well as enhanced superconducting transition temperatures. These observations point to the crucial role played by the critical Ising-nematic fluctuations \cite{oganesyan2001,metzner03,lawler06,sslee09,fradkin10,metlitski10a,mross10,fernandes14,holder15,Karahasanovic2016,Paul2017,lederer15,Berg:2018aa}.

\begin{figure}
  \centering
  \includegraphics[width=0.8\linewidth]{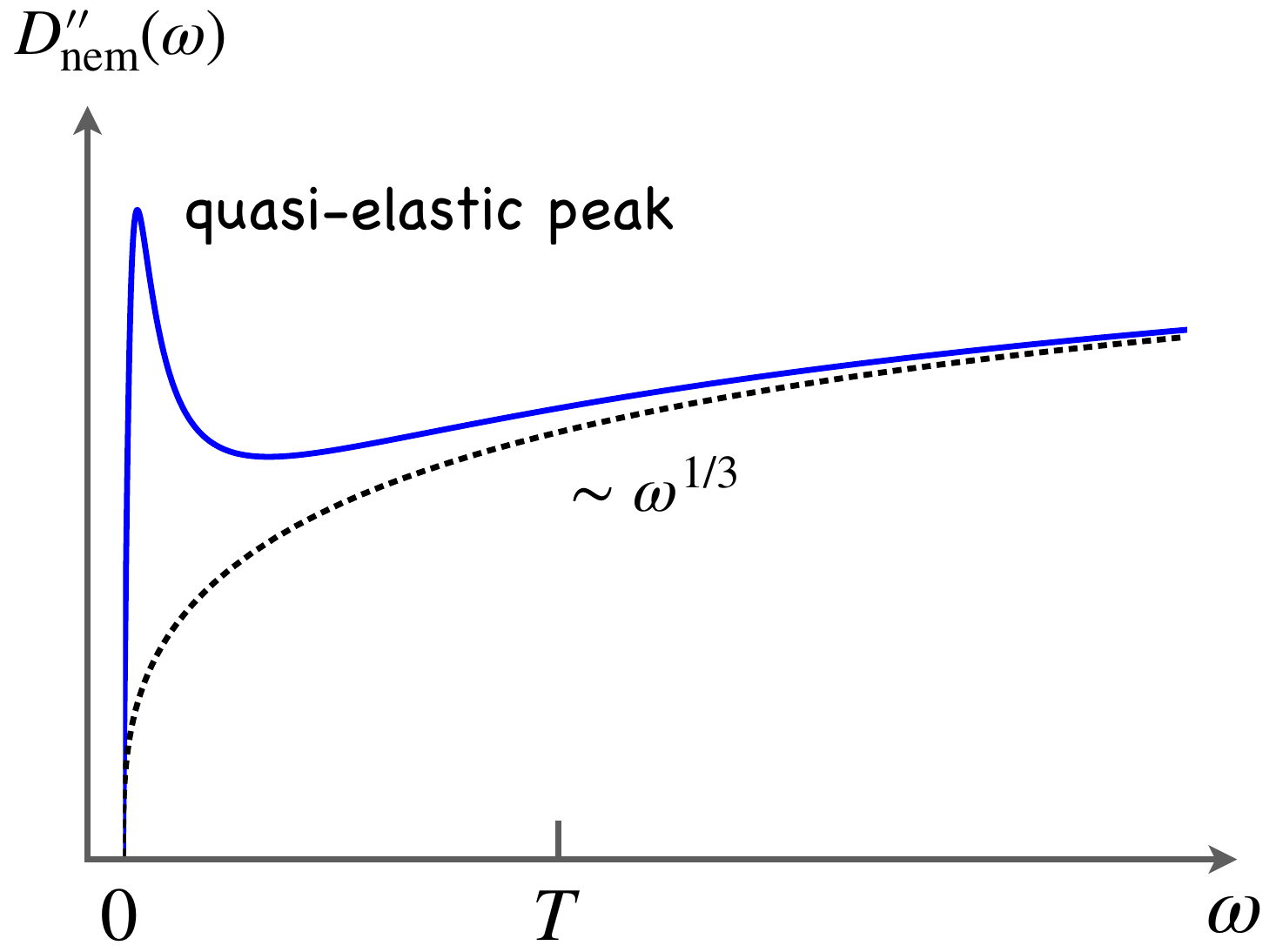}
  \caption{\label{fig:schematic}Schematic plot of the imaginary part of the dynamical nematic susceptibility in the vicinity of an Ising-nematic QCP, featuring a low-frequency quasi-elastic peak, and $\omega^{1/3}$ dependence at frequencies higher than temperature. The Fermi energy $E_F$ is taken to be much larger than temperature.}
\end{figure}

Due to the presence of gapless quasi-particles near the Fermi surface, the dynamical properties of the critical fluctuations are strongly modified compared to those of an insulator. In the quasi-static and long wavelength limit ($\omega \ll |v_F\mathbf{q}|$) the dynamics is governed by ``Landau damping", i.e., the decay of critical fluctuations into collective electron-hole excitations near the Fermi surface. The purely dynamical limit ($\omega \gg |v_F\mathbf{q}|$) is much less studied. Raman scattering experiments in FeSe$_{\text{1-x}}$S$_{\text{x}}$, which probe the dynamics in the latter regime, reveal~\cite{zhang17} a pronounced quasi-elastic peak (QEP, see Fig.~\ref{fig:schematic}) near the Ising-nematic phase transition. The peak height grows proportionally to the thermodynamic nematic susceptibility, and displays a Curie-Weiss behavior as a function of temperature. More interestingly, the peak occurs at a frequency {\it smaller} than temperature, and softens as $T^{\alpha}$ where $\alpha>1$. 
Theoretically, Ref.~\cite{gallais16} showed that a QEP can occur in the presence of impurity scattering, and its frequency is predicted to scale as the inverse thermodynamic susceptibility. 

On the other hand, two recent theoretical works \cite{klein18a,klein18b} studied the nematic dynamical response at zero temperature in the presence electron-electron scattering from critical fluctuations, 
predicting an $\omega^{1/3}$ behavior in the low frequency range. These studies inevitably missed the sub-temperature QEP.


In this Letter, we present a detailed study of the dynamical nematic suscetibility $D_{\text{nem}}(\mathbf{q}\rightarrow 0,\omega)$ at {\it finite temperatures} when a two-dimensional electronic system is driven toward an Ising-nematic QCP. We use a novel memory matrix approach developed recently~\cite{xw2019}, which treats the quasi-particle occupation numbers near the Fermi surface as slow variables~\cite{HQM}. The main result is sketched in Fig.~\ref{fig:schematic}. We argue that the QEP reflects the slow relaxation of an Ising-nematic deformation of the Fermi surface, analogous to the Drude peak in the optical conductivity (which is associated with the slow decay of the current). The  frequency of QEP, $\Gamma(T)$, depends on both dyanmical and thermodynamic properties. We find that $\Gamma(T) \propto \tau^{-1}\chi^{-1}$, where $\chi(T)$ is the Ising-nematic thermodynamic susceptibility, and $\tau(T)$ is the lifetime of the Ising-nematic deformation. This deformation is relaxed by scattering of quasi-particles from the long-wavelength critical fluctuations (with $\omega\ll |v_F\mathbf{q}|$), as well as by impurity scattering. 
The frequency $\Gamma(T)$ 
vanishes at the onset of the Ising-nematic order, where a nematic deformation of the Fermi surface becomes energetically favorable. 
At frequencies higher than the temperature, we find that the dynamical susceptibility scales as $\omega^{1/3}$, consistently with earlier work~\cite{klein18a}.

To set the stage we consider a simple boson-fermion model in two dimensions that realizes an Ising-nematic QCP, given by the action:
\begin{equation}
\begin{split}
    S & = \int_{0}^{1/T}\mathrm{d}\tau \left[ \sum_{\mathbf{k}\sigma}  c^\dagger_{\mathbf{k}\sigma}(\partial_\tau +\varepsilon_\mathbf{k}) c_{\mathbf{k}\sigma}  + \frac{\lambda}{\sqrt{N}} \sum_\mathbf{q} \phi_\mathbf{q} \hat{Q}_{-\mathbf{q}}\right] \\
    & + \int_{0}^{1/T}\mathrm{d}\tau \sum_{\mathbf{q}}\frac{1}{2}\left[ 1+(\mathbf{q}\xi_0)^2 \right]|\phi_\mathbf{q}|^2 .
\end{split} \label{eq:modelH}
\end{equation}
Here $c_{\mathbf{k}\sigma}$ annihilates an electron with momentum $\mathbf{k}$ and spin $\sigma$. $N=2$ is the number of spin components (below, we shall generalize the problem to an arbitrary $N$). For simplicity we consider a parabolic dispersion $\varepsilon_\mathbf{k} = |\mathbf{k}|^2/2m-\mu$.  The bosonic field $\phi_{\mathbf{q}}$ represents the Ising-nematic fluctuations. 
The bare nematic propagator is parametrized by 
the bare correlation length $\xi_0$. 
$\phi$ couples linearly to the fermionic bilinear $\hat{Q}_{-\mathbf{q}} = \sum_{\mathbf{k}\sigma} \varphi_{\mathbf{k},\mathbf{k+q}} c^{\dagger}_{\mathbf{k}\sigma} c_{\mathbf{k+q}\sigma}$, with a coupling strength $\lambda$. $\varphi_{\mathbf{k},\mathbf{k+q}}$ is the Ising-nematic form factor that changes sign under $90$-degree in-plane rotation. 
Note that $\lambda^2$ has unit of energy. We consider an electronically driven nematic QCP due to the coupling term, and set the bare correlation length $\xi_0=k_F^{-1}$, the Fermi wavenumber. This is a strong coupling instability $\lambda_c^2\propto E_F$ analogous to the Stoner instability for ferromagnetism.

The properties of low-energy excitations near the QCP have been studied extensively in the literature, and here we merely quote the results. The long wavelength $\phi$ fluctuations gain dynamics via Landau damping. In the limit $\omega < |v_F \mathbf{q}|\ll E_F$, it is described by the following propagator:
\begin{equation} \label{eq:smallQPropagator}
    D^{-1}(\mathbf{q},\omega) \approx r(T) + (\mathbf{q}\xi_0)^2 - i\gamma_\mathbf{q}\omega,
\end{equation}
where $\gamma_\mathbf{q}\propto \gamma \frac{k_F}{q}\cos 2\theta_\mathbf{q}$ is the Landau-damping coefficient. A one-loop approximation gives $\gamma = \frac{N}{2\pi}\frac{\lambda^2}{v_F^2}$
(the approximation is formally justified over a finite range of energies in the large $N$ limit). Within the same approximation, the renormalized mass satisfies $r(T)\propto T^2$. However, it has been shown from both field theoretical methods~\cite{hartnoll14} and numerical simulations~\cite{schattner16} that $r(T)\propto T$ instead (up to a $\log T$ correction). Throughout this paper, we will assume $r(T)=T$ without a fully self-consistent calculation, while keeping the one-loop form for the Landau damping coefficient.

The feedback of the critical fluctuations on single-electron properties is captured by a self-energy term: $\Sigma(\mathbf{k},\omega)\propto i E_F |\gamma\omega|^{2/3}  \cos^2 2\theta_\mathbf{k}$. The self-energy term becomes dominant in the hot regions below an energy scale $\Omega_\text{NFL} \propto \lambda^4E_F^{-1}N^{-3}$. Below this energy scale, the naive large $N$ approximation breaks down~\cite{sslee09}. The strong dependence on the fermion flavor number suggests that the non-Fermi liquid scale can be parametrically suppressed by going to the large-$N$ limit~\cite{mross15,xw2019}. This will be assumed to be true throughout this paper, and the physics below the non-Fermi liquid scale is left to future studies.

In the coherent electron regime ($\Omega_{\mathrm{NFL}}\ll T\ll E_F$) the transport properties can be described using a kinetic equation approach, where the effects of critical fluctuations are incorporated into Fermi liquid parameters and the collision integral. In an earlier work \cite{xw2019}, we have shown that the kinetic equation can be derived microscopically using the memory matrix formalism \cite{HQM,forster}, treating the electron occupation number in the momentum space $\{\hat{n}_{\mathbf{k}\sigma}\equiv c^\dagger_{\mathbf{k}\sigma}c_{\mathbf{k}\sigma}\}$ as the subspace of ``slow" operators.

The dynamical nematic susceptibility is defined via:
\begin{equation}\label{eq:RamanDef}
    D_{\text{nem}}(\omega) = i \int_{0}^{\infty}\mathrm{d}t\ \exp(i\omega t)\langle \left[\hat{Q}(t),\hat{Q}(0)\right]\rangle,
\end{equation}
where $\hat{Q}\equiv\hat{Q}_{\mathbf{q}=0}$. It is straightforward to show (See Ref.~\cite{xw2019}) that Eq.~(\ref{eq:RamanDef}) can be casted into the following memory matrix expression:
\begin{equation}
\begin{split}
    & D_{\text{nem}}(\omega) \approx  \chi_{\hat{Q},\hat{Q}} \\
    & + \sum_{\mathbf{k}\sigma;\mathbf{k'}\sigma'}  \chi_{\hat{Q},\hat{n}_{\mathbf{k}\sigma}} \left[\frac{i\omega}{M(\omega)-i\omega \chi}\right]_{\hat{n}_{\mathbf{k}\sigma},\hat{n}_{\mathbf{k}'\sigma'}}\chi_{\hat{n}_{\mathbf{k}'\sigma'},\hat{Q}}
\end{split} \label{eq:RamanB1g}.
\end{equation}
Here $\chi_{A,B}\equiv \int_0^{1/T}\mathrm{d}\tau \left[ \langle A(\tau)B(0)\rangle - \langle A\rangle \langle B\rangle \right]$. 
The memory matrix $M$ is defined as:
\begin{equation}
    M_{\hat{n}_{\mathbf{k}\sigma},\hat{n}_{\mathbf{k'}\sigma'}}(\omega) = \frac{1}{i\omega}\left[G^R_{\dot{\hat{n}}_{\mathbf{k}\sigma},\dot{\hat{n}}_{\mathbf{k'}\sigma'}}(\omega) - G^R_{\dot{\hat{n}}_{\mathbf{k}\sigma},\dot{\hat{n}}_{\mathbf{k'}\sigma'}}(0)\right], \label{eq:mmdef}
\end{equation}
where $G^R_{A,B}(t)\equiv -i\Theta(t)\langle [A(t),B(0)]\rangle$ is the retarded Green's function. We have defined $\dot{\hat{n}}_{\mathbf{k}\sigma}=i[H,\hat{n}_{\mathbf{k}\sigma}]$, where $H$ is the Hamiltonian that corresponds to the action~\eqref{eq:modelH}. 
In our system, $\dot{\hat{n}}_{\mathbf{k}\sigma}$ is given by:
\begin{equation}
    \dot{\hat{n}}_{\mathbf{k}\sigma}= i\lambda \sum_{\mathbf{q}} \phi_\mathbf{q}\left( \varphi_{\mathbf{k},\mathbf{k-q}}c^\dagger_{\mathbf{k}\sigma}c_{\mathbf{k-q}\sigma}-\varphi_{\mathbf{k},\mathbf{k+q}}c^\dagger_{\mathbf{k+q}\sigma}c_{\mathbf{k}\sigma}\right).
\end{equation}

The connection of Eq.~(\ref{eq:RamanB1g}) to a kinetic equation is made by identifying the memory matrix as the linearized collision integral, and the thermodynamic susceptibilities as the Fermi liquid parameters. To see that $D_{\mathrm{nem}}(\omega)$ contains a QEP, we treat for simplicity $\hat{Q}$ as the only slow operator (a more rigorous treatment will follow). We arrive at a {\it memory function} expression:
\begin{equation} \label{eq:RamanQualitative}
  D_\text{nem}(\omega) \approx  \frac{M_{\hat{Q},\hat{Q}}(\omega) \chi_{\hat{Q},\hat{Q}}}{M_{\hat{Q},\hat{Q}}(\omega)-i\omega \chi_{\hat{Q},\hat{Q}}}.
\end{equation}
We write the memory function as $M_{\hat{Q},\hat{Q}}(\omega,T)=M'(\omega,T)+i M''(\omega,T)$. At low frequencies and finite temperatures, the real part can be approximated by its $\omega\rightarrow 0$ value, which 
vanishes as a power law of temperature \cite{xw2019}. The imaginary part is approximately linear in frequency, and renormalizes the strength of the two-particle response. This renormalization is non-singular and subleading in $1/N$, and hence we neglect it. As a result, $\mathrm{Im}D_{\mathrm{nem}}(\omega)$ contains a low-frequency peak at $\omega \approx M'_{\hat{Q},\hat{Q}}(\omega=0)/\chi_{\hat{Q},\hat{Q}}$, with a peak height of $\chi_{\hat{Q},\hat{Q}}/2$.

\begin{figure}
    \centering
    \includegraphics[width=0.95\linewidth]{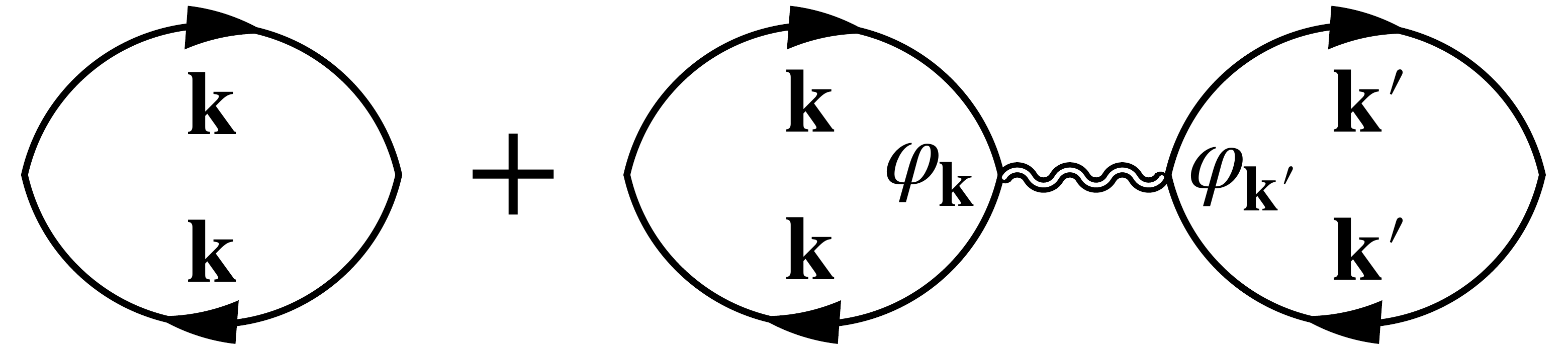}
    \caption{Feynman diagrams for the thermodynamic susceptibility $\chi_{\hat{n}_\mathbf{k},\hat{n}_\mathbf{k'}}$. The double-wiggly line represents the dressed propagator for nematic fluctuations.}
    \label{fig:feynman}
\end{figure}

We perform a diagrammatic calculation of both the memory matrix and the thermodynamic susceptibility to leading order in $1/N$, equivalent to the random phase approximation (RPA). The leading order Feynman diagrams for the thermodynamic susceptibilities are shown in Fig.~\ref{fig:feynman}, where the double-curly line represent the dressed bosonic propagator in Eq.~(\ref{eq:smallQPropagator}), and we have introduced a short-hand notation: $\varphi_\mathbf{k} \equiv \varphi_{\mathbf{k},\mathbf{k}}$. We consider the limit when temperature is much smaller than the Fermi energy, and hereby work with the following approximation:
\begin{equation}
   \chi_{\hat{n}_{\mathbf{k}\sigma},\hat{n}_{\mathbf{k}'\sigma'}} \approx   \delta_{\mathbf{k}\sigma,\mathbf{k}'\sigma'} \delta(\varepsilon_{\mathbf{k}}) + \frac{\lambda^2}{r(T)}\varphi_\mathbf{k}\varphi_{\mathbf{k}'} \delta(\varepsilon_\mathbf{k})\delta(\varepsilon_{\mathbf{k}'}).
\end{equation}
The presence of the $\delta$-functions indicates that the main contribution comes from the vicinity of the Fermi surface.

The diagrammatic calculation of the memory matrix (Eq.~(\ref{eq:mmdef})) has been discussed in detail in Ref.~\cite{xw2019}. Essentially, the RPA for the memory matrix is a conserving approximation, where the conservation of total electronic number and momentum are built into the formalism. Moreover, the high-frequency expansion of Eq.~(\ref{eq:RamanB1g}) has been shown to reproduce the perturbative results when both the Maki-Thompson and Aslamazov-Larkin diagrams are considered.

Below we consider two limiting cases where either $\omega \ll T \ll E_F$ (quasi-elastic limit) or $T\ll \omega \ll E_F$ (intermediate frequencies).

\textit{Quasi-elastic limit.--}
In this limit, it is convenient to work in the angular momentum basis $M_{n\sigma,m\sigma'}= \sum_{\mathbf{k}\mathbf{k}'} e^{-i(m\theta_\mathbf{k}-n\theta_{\mathbf{k}'})}M_{\mathbf{k}\sigma,\mathbf{k}'\sigma'}$, where the memory matrix has a simple form ~\cite{xw2019}:
\begin{equation} \label{eq:mm_dc}
  \begin{split}
      & M_{n\sigma,m\sigma'}(\omega\approx 0,T)  \approx \pi \lambda^2 \sum_{\mathbf{k}\mathbf{k}'}f^{*}_{n,\mathbf{k}\mathbf{k}'}f_{m,\mathbf{k}\mathbf{k}'}V_{\mathbf{k}- \mathbf{k}'}(T) \\
      & \times \varphi_{\mathbf{k},\mathbf{k}'}^2\delta(\varepsilon_\mathbf{k})\delta(\varepsilon_{\mathbf{k}'}) \left[1-\frac{1-(-1)^n}{2}\frac{1-(-1)^m}{2}\right]\delta_{\sigma\sigma'},
  \end{split}
\end{equation}
where $f_{m,\mathbf{k}\mathbf{k}'}= \exp(i m \theta_\mathbf{k})-\exp(i m \theta_{\mathbf{k}'})$, and
\begin{equation} \label{eq:Vq}
  V_\mathbf{q}(T) = \int_{-\infty}^{\infty}\frac{\mathrm{d}\omega}{\pi} \omega D''(\mathbf{q},\omega)\left(-\frac{\partial n_B(\omega)}{\partial \omega }\right).
\end{equation}

According to Eq.~(\ref{eq:mm_dc}) the even and odd parity harmonics behave differently. For odd parity modes (such as total electronic momentum), the terms in the brackets vanish. Hence, odd parity defonrmations of the FS are conserved within the approximation that all scattering processes occur at the Fermi surface. This is unique to two-dimensional electronic systems, as has been discussed in Ref.~\cite{maslov11,ledwidth17}. On the other hand, even parity modes such as $\hat{Q}$ have a finite lifetime. Furthermore, different even-parity modes are coupled due to the angular form factor of the critical fluctuations.

In the angular harmonics basis, Eq.~(\ref{eq:RamanB1g}) has a simpler expression, given by:
\begin{equation} \label{eq:Ramandc}
  D_\text{nem}(\omega) \approx  \chi_{\hat{Q},\hat{Q}} + i\omega \chi_{\hat{Q},\hat{Q}}^2 \left[ \frac{1}{M(\omega)-i\omega \chi} \right]_{\hat{Q},\hat{Q}}.
\end{equation}
Note that there is a crucial difference compared to Eq.~(\ref{eq:RamanQualitative}). Here the matrix in the square brackets is first inverted before taking the overlap with the Ising-nematic form factor. Due to the hybridization of different angular harmonics [see Eq.~(\ref{eq:mm_dc})], the two expressions may give qualitatively different results.

We first present a qualitative analysis for the temperature dependence of the QEP frequency. Since critical fluctuations give rise to small-angle scattering, we expect that the decay of an Ising-nematic deformation of the Fermi surface to be governed by momentum diffusion, which leads to a decay rate $\tau^{-1}\propto (q_0/k_F)^2 \tau^{-1}_{0}$, where $\tau_0^{-1}$ is the single-particle scattering rate, and $q_0$ is the characteristic momentum transfer. $q_0\sim T^{1/3}$ and $\tau^{-1}_{0}\sim T^{2/3}$ following quantum critical scaling. As a result, the QEP frequency should scale as $\tau^{-1}\chi^{-1}\propto T^{7/3}$ near the QCP.

\begin{figure}
    \centering
    \includegraphics[width=\linewidth]{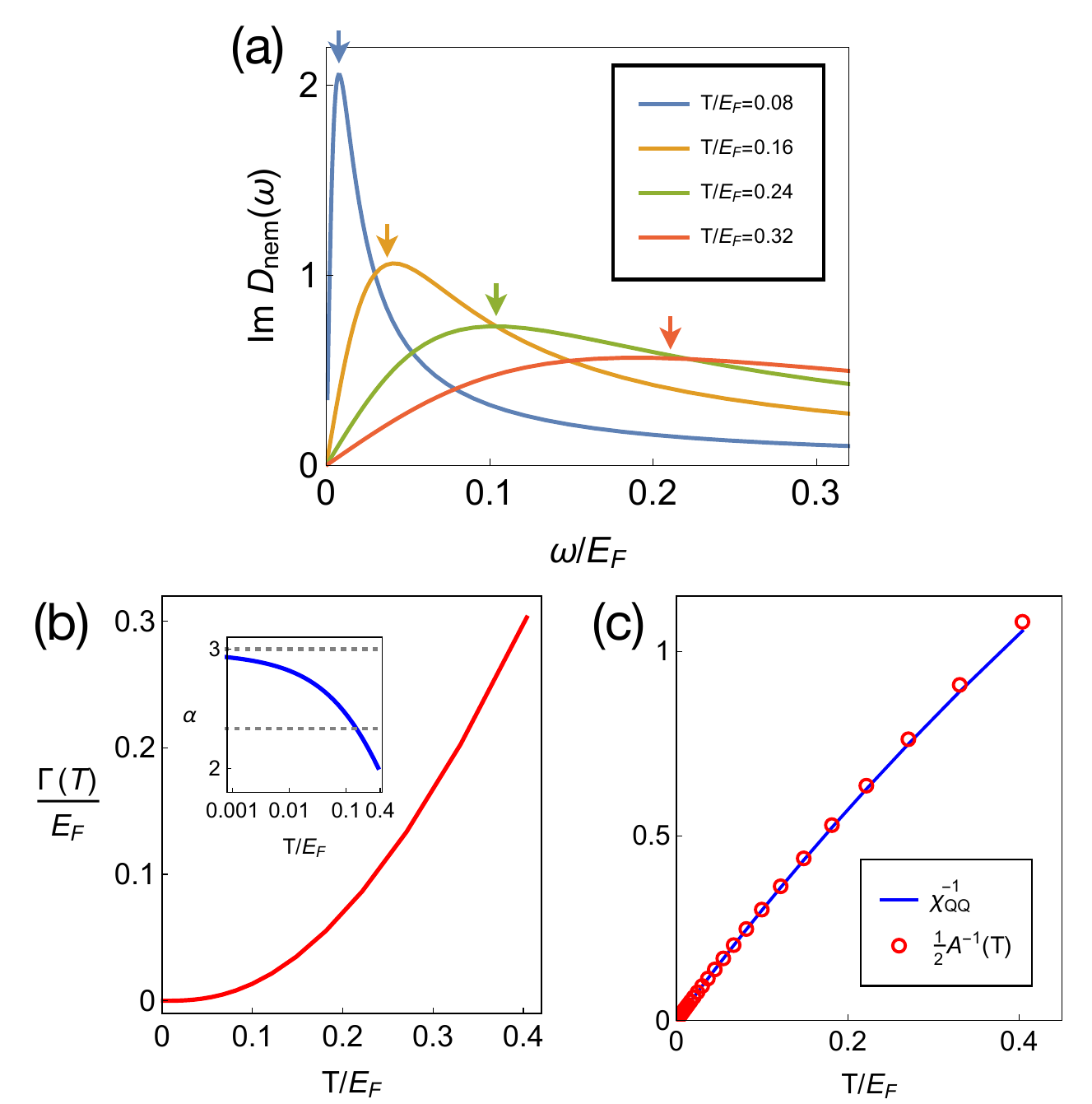}
    \caption{ (a) Imaginary part of the dynamical nematic susceptibility showing a quasi-elastic peak feature for various temperatures. (b) QEP peak frequency $\Gamma(T)$ as a function of temperature. Inset is the log-derivative plot $\alpha \equiv d\ln \Gamma/d\ln T$ showing the temperature variation of the power law exponent. The dashed lines correspond to a $T^2\chi^{-1}$ (Fermi liquid behavior) and $T^{4/3}\chi^{-1}$.
    (c) QEP peak height $A(T)$ (red circle) compared the thermodynamic susceptibility (blue solid line).}
    \label{fig:numerical_results}
\end{figure}

In Fig.~\ref{fig:numerical_results} we present a numerical solution for the low-frequency Raman response for a clean system, replacing the memory matrix by its $\omega \rightarrow 0$ limit. We choose 
$\lambda^2 = 4\pi E_F$ (close to the RPA instability). Fig.~\ref{fig:numerical_results}(a) shows the response for various temperatures. The spectral response clearly shows the development of a low-frequency peak as temperature is lowered toward the QCP. The temperature dependence of the peak height and peak frequency are presented in Fig.~\ref{fig:numerical_results}(b) and (c) respectively. As expected, the peak height tracks the thermodynamic susceptibility, while the peak frequency softens toward the QCP. However, the temperature dependence of the peak frequency cannot be fitted to a simple power law governed by momentum diffusion. The log-derivative plot (Fig.~\ref{fig:numerical_results}(b) inset) shows a smooth variation of the exponent as temperature is lowered, saturating to $T^3$ at low temperatures. This implies that $\tau^{-1}\propto T^2$ -- analogous to that of a Fermi liquid, even though we are at the QCP.

The apparent violation of the naive quantum critical scaling at all temperatures can be understood as follows. At high temperatures, $V_\mathbf{q}(T)\propto \frac{T\gamma_\mathbf{q}}{r(T)+\mathbf{q}^2}$ from Eq.~(\ref{eq:Vq}). The typical momentum transfer $q_0\sim k_F\ll r(T)$. As a result, $V_\mathbf{q}(T)\propto \frac{T\gamma_{q_0}}{r(T)} \sim \text{const}$, leading to a constant lifetime. At low temperatures, 
the typical momentum transfer $q_0\sim k_F(\gamma T)^{1/3}$ is small. Furthermore, the Fermi surface is divided into four weakly-connected patches by the Ising-nematic cold spots, where the form factor vanish by symmetry. Scattering across the cold spots is the bottleneck for global equilibration. Evaluating the memory matrix (Eq.~(\ref{eq:mm_dc})) near the cold spots, we get $M\propto \int_0^{q_0}\mathrm{d}^2q q^4V_q \sim T^2$. Compared to rate of momentum diffusion, there is an additional factor of $q_0^2$ due to the form factor.

\begin{figure}
    \centering
    \includegraphics[width=\linewidth]{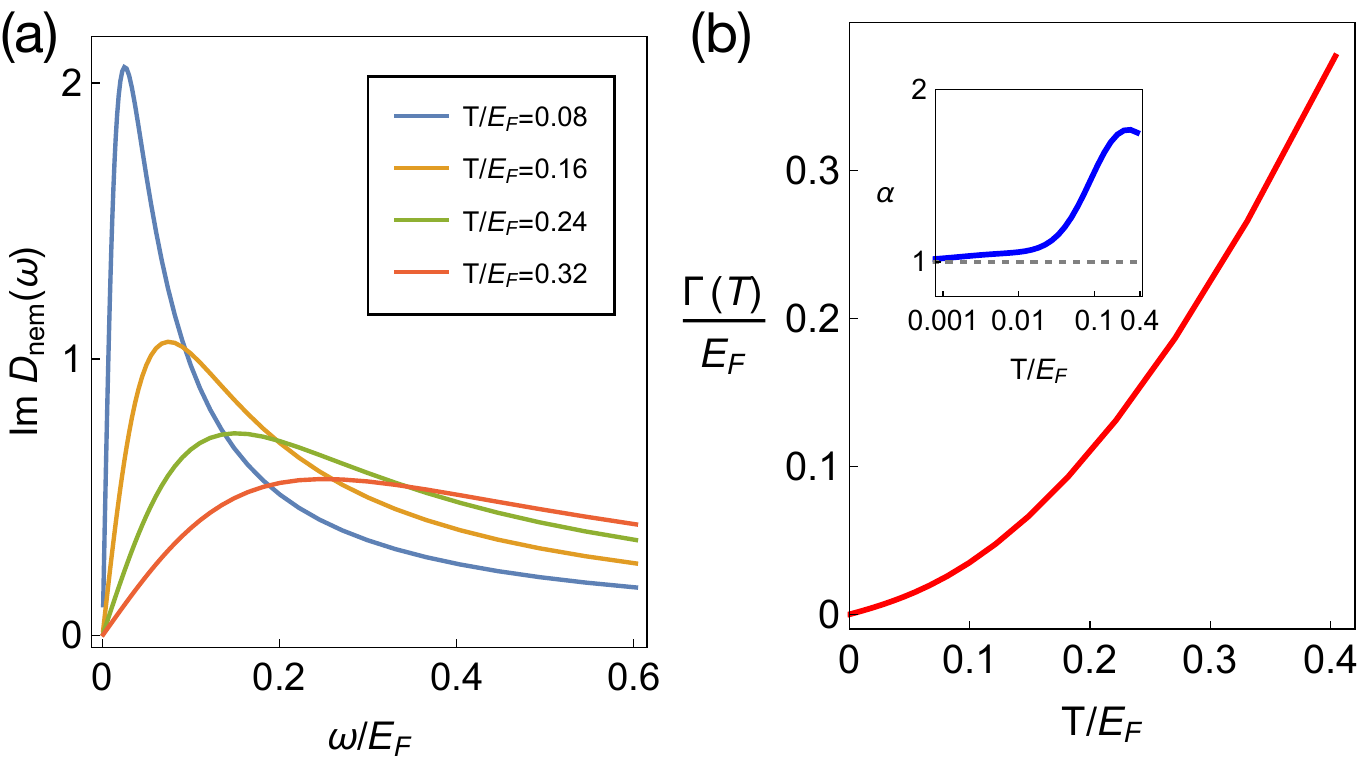}
    \caption{(a) Imaginary part of the dynamical nematic susceptibility in the presence of impurity scattering ($g_{\text{imp}}=0.1E_F$). (b) QEP peak frequency $\Gamma(T)$ as a function of temperature. Inset is the log-derivative plot showing the temperature variation of the power law exponent $\alpha$.}
    \label{fig:numerical_impurity}
\end{figure}

Next, we consider how impurity scattering affects the properties of the QEP. For simplicity we assume the impurity scattering only contributes to the memory matrix $M(T) = M_{\text{nem}}(T) + M_{\text{imp}}(T)$ but does not modify the thermodynamic properties. Now, the cold spots are no longer a bottlneck for relaxation, and as a result $\Gamma(T)\propto \left( \tau_\text{imp}^{-1} + \tau_{\text{nem}}^{-1}(T) \right) \chi^{-1}$. At low temperatures impurity scattering is dominant, leading to $\Gamma(T)\propto T$ coming from the thermodynamic susceptibility. This is the behavior discussed in Ref.~\cite{gallais16}. At higher temperatures, nematic fluctuations are more important, leading to a stronger temperature dependence. This behavior is illustrated in Fig.~\ref{fig:numerical_impurity}.

\textit{Intermediate frequencies.--} We proceed to study the limit where the external frequency is much larger than temperature. Here, 
the memory matrix in Eq.~(\ref{eq:RamanB1g}) is always smaller than $\omega \chi_{\hat{Q},\hat{Q}}$. As a result, the approximate memory function expression in Eq.~(\ref{eq:RamanQualitative}) holds. It is straightforward to show that $M_{\hat{Q},\hat{Q}}(\omega)\propto \omega^{4/3}$ governed by momentum diffusion. The imaginary part of the Raman response is then given by:
\begin{equation}
  \mathrm{Im}D_\text{nem}(\omega) \approx \frac{M_\text{Q,Q}(\omega)}{\omega} \propto \omega^{1/3}.
\end{equation}
This behavior has also been obtained in earlier works using perturbative diagrammatic techniques at zero temperature \cite{klein18a,klein18b}.

So far we have neglected the effects of acoustic phonons. As argued in Ref.~\cite{Paul2017}, the nemato-elastic coupling shifts the position of the Ising-nematic phase transition temperature. Experimentally this is reflected in the difference between the extrapolated Curie-Weiss temperature $T_\Theta$ and measured transition temperature $T_s$~\cite{zhang17}. Moreover, the coupling leads to 
``directional criticality" at the QCP, where the correlation length is divergent only along the diagonal directions of the Brillouin zone. As discussed in Refs.~\cite{Karahasanovic2016,Paul2017,carvalho2019a,carvalho2019b}, this can also lead to a breakdown of quantum critical scaling, and recover Fermi liquid behavior at low temperatures.

\begin{figure}
    \centering
    \includegraphics[width=\linewidth]{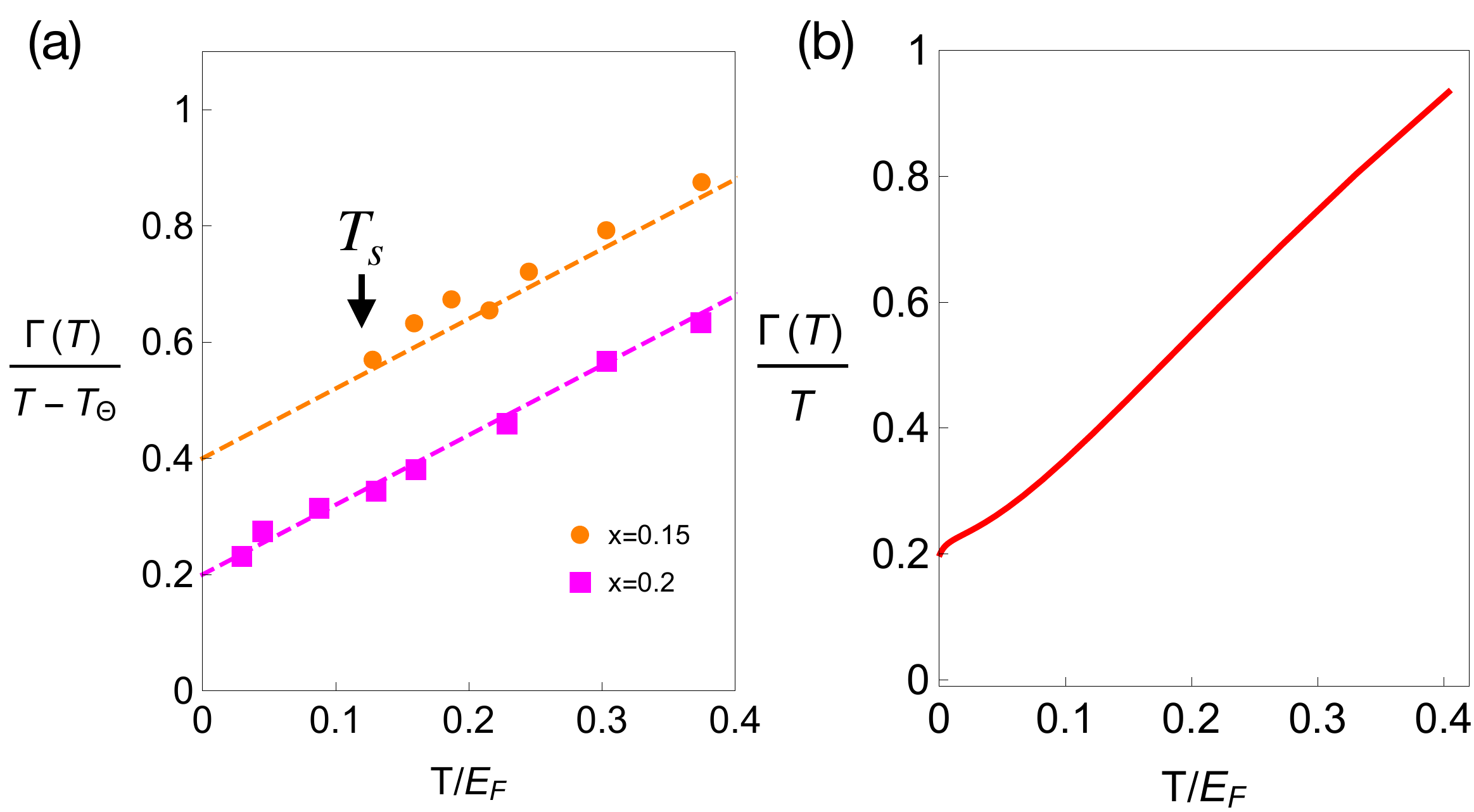}
    \caption{(a) Quasi-elastic peak frequency for FeSe$_\text{1-x}$S$_\text{x}$ at dopings $x=0.15$ (orange) and $x=0.2$ (magenta), extracted from Ref.~\cite{zhang17}. The extrapolated Curie-Weiss temperature is zero at $x=0.15$, and negative at $x=0.2$. $E_F$ is taken to be 30meV. The dashed lines are linear fittings using $a+1.2T/E_F$, with $a=0.2$ and $0.4$ respectively. (b) A re-plot of Fig.~\ref{fig:numerical_impurity}(b) with the y-axis being $\Gamma(T)/T$.}
    \label{fig:comparisons}
\end{figure}

We proceed to compare our results to Raman scattering experiments performed on FeSe$_\text{1-x}$S$_\text{x}$~\cite{zhang17}, where a putative Ising-nematic QCP occurs at $x\approx 0.15$. In Fig.~\ref{fig:comparisons}(a) we plot $\Gamma(T)/(T-T_\Theta)$ as a function of $T/E_F$ for $x=0.15,0.2$, extracted from the experiments. As discussed earlier, this quantity represents the dynamical contributions to the QEP. Above $T_s(x)$, the data can be fitted using a functional form $a+b T$, suggesting that the near-critical Ising-nematic fluctuations give rise to a linear-in-T scattering rate. This behavior is qualitatively captured from our calculation shown in Fig.~\ref{fig:comparisons}(b) --- a replot of Fig.~\ref{fig:numerical_impurity}(b), without fine-tuning of parameters except the strength of impurity scattering.

In summary, using a memory matrix approach, we studied the dynamical nematic susceptibility for a two-dimensional electronic system near an Ising-nematic quantum critical point. Our results are qualitatively consistent with the Raman scattering results for FeSe$_\text{1-x}$S$_\text{x}$. Interestingly, we find that the quasi-elastic peak frequency $\Gamma(T)\propto \tau^{-1}\chi^{-1}$. As a result, a dynamical scattering rate $1/\tau$ can be extracted directly from experimental data, by dividing out the thermodynamic susceptibility. 

\textit{Acknowledgement} We acknowledge fruitful discussions with Girsh Blumberg, Andrey Chubukov, and Avraham Klein. XW acknowledge financial support from National MagLab, which is funded by the National Science Foundation (DMR-1644779) and the state of Florida. EB acknowledges support from the European Research Council (ERC) under grant HQMAT (Grant Agreement No. 817799), the Israel-USA Binational Science Foundation (BSF) grant no. 2018217, and the Minerva foundation.  

\bibliographystyle{apsrev4-1}
\bibliography{references}

\begin{thebibliography}{36}%
\makeatletter
\providecommand \@ifxundefined [1]{%
 \@ifx{#1\undefined}
}%
\providecommand \@ifnum [1]{%
 \ifnum #1\expandafter \@firstoftwo
 \else \expandafter \@secondoftwo
 \fi
}%
\providecommand \@ifx [1]{%
 \ifx #1\expandafter \@firstoftwo
 \else \expandafter \@secondoftwo
 \fi
}%
\providecommand \natexlab [1]{#1}%
\providecommand \enquote  [1]{``#1''}%
\providecommand \bibnamefont  [1]{#1}%
\providecommand \bibfnamefont [1]{#1}%
\providecommand \citenamefont [1]{#1}%
\providecommand \href@noop [0]{\@secondoftwo}%
\providecommand \href [0]{\begingroup \@sanitize@url \@href}%
\providecommand \@href[1]{\@@startlink{#1}\@@href}%
\providecommand \@@href[1]{\endgroup#1\@@endlink}%
\providecommand \@sanitize@url [0]{\catcode `\\12\catcode `\$12\catcode
  `\&12\catcode `\#12\catcode `\^12\catcode `\_12\catcode `\%12\relax}%
\providecommand \@@startlink[1]{}%
\providecommand \@@endlink[0]{}%
\providecommand \url  [0]{\begingroup\@sanitize@url \@url }%
\providecommand \@url [1]{\endgroup\@href {#1}{\urlprefix }}%
\providecommand \urlprefix  [0]{URL }%
\providecommand \Eprint [0]{\href }%
\providecommand \doibase [0]{http://dx.doi.org/}%
\providecommand \selectlanguage [0]{\@gobble}%
\providecommand \bibinfo  [0]{\@secondoftwo}%
\providecommand \bibfield  [0]{\@secondoftwo}%
\providecommand \translation [1]{[#1]}%
\providecommand \BibitemOpen [0]{}%
\providecommand \bibitemStop [0]{}%
\providecommand \bibitemNoStop [0]{.\EOS\space}%
\providecommand \EOS [0]{\spacefactor3000\relax}%
\providecommand \BibitemShut  [1]{\csname bibitem#1\endcsname}%
\let\auto@bib@innerbib\@empty
\bibitem [{\citenamefont {Zhang}\ \emph {et~al.}(2017)\citenamefont {Zhang},
  \citenamefont {Wu}, \citenamefont {Kasahara}, \citenamefont {Shibauchi},
  \citenamefont {Matsuda},\ and\ \citenamefont {Blumberg}}]{zhang17}%
  \BibitemOpen
  \bibfield  {author} {\bibinfo {author} {\bibfnamefont {W.~L.}\ \bibnamefont
  {Zhang}}, \bibinfo {author} {\bibfnamefont {S.~F.}\ \bibnamefont {Wu}},
  \bibinfo {author} {\bibfnamefont {S.}~\bibnamefont {Kasahara}}, \bibinfo
  {author} {\bibfnamefont {T.}~\bibnamefont {Shibauchi}}, \bibinfo {author}
  {\bibfnamefont {Y.}~\bibnamefont {Matsuda}}, \ and\ \bibinfo {author}
  {\bibfnamefont {G.}~\bibnamefont {Blumberg}},\ }\href@noop {} {\enquote
  {\bibinfo {title} {{Stripe quadrupole order in the nematic phase of
  FeSe$_{1-x}$S$_x$}},}\ } (\bibinfo {year} {2017}),\ \Eprint
  {http://arxiv.org/abs/arXiv:1710.09892} {arXiv:1710.09892} \BibitemShut
  {NoStop}%
\bibitem [{\citenamefont {Klein}\ \emph
  {et~al.}(2018{\natexlab{a}})\citenamefont {Klein}, \citenamefont {Lederer},
  \citenamefont {Chowdhury}, \citenamefont {Berg},\ and\ \citenamefont
  {Chubukov}}]{klein18a}%
  \BibitemOpen
  \bibfield  {author} {\bibinfo {author} {\bibfnamefont {A.}~\bibnamefont
  {Klein}}, \bibinfo {author} {\bibfnamefont {S.}~\bibnamefont {Lederer}},
  \bibinfo {author} {\bibfnamefont {D.}~\bibnamefont {Chowdhury}}, \bibinfo
  {author} {\bibfnamefont {E.}~\bibnamefont {Berg}}, \ and\ \bibinfo {author}
  {\bibfnamefont {A.}~\bibnamefont {Chubukov}},\ }\href {\doibase
  10.1103/PhysRevB.98.041101} {\bibfield  {journal} {\bibinfo  {journal} {Phys.
  Rev. B}\ }\textbf {\bibinfo {volume} {98}},\ \bibinfo {pages} {041101}
  (\bibinfo {year} {2018}{\natexlab{a}})}\BibitemShut {NoStop}%
\bibitem [{\citenamefont {Daou}\ \emph {et~al.}(2010)\citenamefont {Daou},
  \citenamefont {Chang}, \citenamefont {LeBoeuf}, \citenamefont
  {Cyr-Choini{\`e}re}, \citenamefont {Lalibert{\'e}}, \citenamefont
  {Doiron-Leyraud}, \citenamefont {Ramshaw}, \citenamefont {Liang},
  \citenamefont {Bonn}, \citenamefont {Hardy},\ and\ \citenamefont
  {et~al.}}]{Daou_2010}%
  \BibitemOpen
  \bibfield  {author} {\bibinfo {author} {\bibfnamefont {R.}~\bibnamefont
  {Daou}}, \bibinfo {author} {\bibfnamefont {J.}~\bibnamefont {Chang}},
  \bibinfo {author} {\bibfnamefont {D.}~\bibnamefont {LeBoeuf}}, \bibinfo
  {author} {\bibfnamefont {O.}~\bibnamefont {Cyr-Choini{\`e}re}}, \bibinfo
  {author} {\bibfnamefont {F.}~\bibnamefont {Lalibert{\'e}}}, \bibinfo {author}
  {\bibfnamefont {N.}~\bibnamefont {Doiron-Leyraud}}, \bibinfo {author}
  {\bibfnamefont {B.~J.}\ \bibnamefont {Ramshaw}}, \bibinfo {author}
  {\bibfnamefont {R.}~\bibnamefont {Liang}}, \bibinfo {author} {\bibfnamefont
  {D.~A.}\ \bibnamefont {Bonn}}, \bibinfo {author} {\bibfnamefont {W.~N.}\
  \bibnamefont {Hardy}}, \ and\ \bibinfo {author} {\bibnamefont {et~al.}},\
  }\href {\doibase 10.1038/nature08716} {\bibfield  {journal} {\bibinfo
  {journal} {Nature}\ }\textbf {\bibinfo {volume} {463}},\ \bibinfo {pages}
  {519} (\bibinfo {year} {2010})}\BibitemShut {NoStop}%
\bibitem [{\citenamefont {Chu}\ \emph {et~al.}(2010)\citenamefont {Chu},
  \citenamefont {Analytis}, \citenamefont {De~Greve}, \citenamefont {McMahon},
  \citenamefont {Islam}, \citenamefont {Yamamoto},\ and\ \citenamefont
  {Fisher}}]{Chu:2010aa}%
  \BibitemOpen
  \bibfield  {author} {\bibinfo {author} {\bibfnamefont {J.-H.}\ \bibnamefont
  {Chu}}, \bibinfo {author} {\bibfnamefont {J.~G.}\ \bibnamefont {Analytis}},
  \bibinfo {author} {\bibfnamefont {K.}~\bibnamefont {De~Greve}}, \bibinfo
  {author} {\bibfnamefont {P.~L.}\ \bibnamefont {McMahon}}, \bibinfo {author}
  {\bibfnamefont {Z.}~\bibnamefont {Islam}}, \bibinfo {author} {\bibfnamefont
  {Y.}~\bibnamefont {Yamamoto}}, \ and\ \bibinfo {author} {\bibfnamefont
  {I.~R.}\ \bibnamefont {Fisher}},\ }\href@noop {} {\bibfield  {journal}
  {\bibinfo  {journal} {Science}\ }\textbf {\bibinfo {volume} {329}},\ \bibinfo
  {pages} {824} (\bibinfo {year} {2010})}\BibitemShut {NoStop}%
\bibitem [{\citenamefont {Kohsaka}\ \emph {et~al.}(2007)\citenamefont
  {Kohsaka}, \citenamefont {Taylor}, \citenamefont {Fujita}, \citenamefont
  {Schmidt}, \citenamefont {Lupien}, \citenamefont {Hanaguri}, \citenamefont
  {Azuma}, \citenamefont {Takano}, \citenamefont {Eisaki}, \citenamefont
  {Takagi}, \citenamefont {Uchida},\ and\ \citenamefont
  {Davis}}]{Kohsaka:2007aa}%
  \BibitemOpen
  \bibfield  {author} {\bibinfo {author} {\bibfnamefont {Y.}~\bibnamefont
  {Kohsaka}}, \bibinfo {author} {\bibfnamefont {C.}~\bibnamefont {Taylor}},
  \bibinfo {author} {\bibfnamefont {K.}~\bibnamefont {Fujita}}, \bibinfo
  {author} {\bibfnamefont {A.}~\bibnamefont {Schmidt}}, \bibinfo {author}
  {\bibfnamefont {C.}~\bibnamefont {Lupien}}, \bibinfo {author} {\bibfnamefont
  {T.}~\bibnamefont {Hanaguri}}, \bibinfo {author} {\bibfnamefont
  {M.}~\bibnamefont {Azuma}}, \bibinfo {author} {\bibfnamefont
  {M.}~\bibnamefont {Takano}}, \bibinfo {author} {\bibfnamefont
  {H.}~\bibnamefont {Eisaki}}, \bibinfo {author} {\bibfnamefont
  {H.}~\bibnamefont {Takagi}}, \bibinfo {author} {\bibfnamefont
  {S.}~\bibnamefont {Uchida}}, \ and\ \bibinfo {author} {\bibfnamefont
  {J.}~\bibnamefont {Davis}},\ }\href@noop {} {\bibfield  {journal} {\bibinfo
  {journal} {Science}\ }\textbf {\bibinfo {volume} {315}},\ \bibinfo {pages}
  {1380} (\bibinfo {year} {2007})}\BibitemShut {NoStop}%
\bibitem [{\citenamefont {B{\"o}hmer}\ and\ \citenamefont
  {Meingast}(2016)}]{bohmer16}%
  \BibitemOpen
  \bibfield  {author} {\bibinfo {author} {\bibfnamefont {A.~E.}\ \bibnamefont
  {B{\"o}hmer}}\ and\ \bibinfo {author} {\bibfnamefont {C.}~\bibnamefont
  {Meingast}},\ }\href
  {http://www.sciencedirect.com/science/article/pii/S1631070515001279}
  {\bibfield  {journal} {\bibinfo  {journal} {Comptes Rendus Physique}\
  }\textbf {\bibinfo {volume} {17}},\ \bibinfo {pages} {90 } (\bibinfo {year}
  {2016})}\BibitemShut {NoStop}%
\bibitem [{\citenamefont {Coldea}\ and\ \citenamefont
  {Watson}(2018)}]{coldea18}%
  \BibitemOpen
  \bibfield  {author} {\bibinfo {author} {\bibfnamefont {A.~I.}\ \bibnamefont
  {Coldea}}\ and\ \bibinfo {author} {\bibfnamefont {M.~D.}\ \bibnamefont
  {Watson}},\ }\href@noop {} {\bibfield  {journal} {\bibinfo  {journal} {Annual
  Review of Condensed Matter Physics}\ }\textbf {\bibinfo {volume} {9}},\
  \bibinfo {pages} {125} (\bibinfo {year} {2018})}\BibitemShut {NoStop}%
\bibitem [{\citenamefont {Reiss}\ \emph {et~al.}()\citenamefont {Reiss},
  \citenamefont {Graf}, \citenamefont {Haghighirad}, \citenamefont {Knafo},
  \citenamefont {Drigo}, \citenamefont {Bristow}, \citenamefont {Schofield},\
  and\ \citenamefont {Coldea}}]{reiss18}%
  \BibitemOpen
  \bibfield  {author} {\bibinfo {author} {\bibfnamefont {P.}~\bibnamefont
  {Reiss}}, \bibinfo {author} {\bibfnamefont {D.}~\bibnamefont {Graf}},
  \bibinfo {author} {\bibfnamefont {A.~A.}\ \bibnamefont {Haghighirad}},
  \bibinfo {author} {\bibfnamefont {W.}~\bibnamefont {Knafo}}, \bibinfo
  {author} {\bibfnamefont {L.}~\bibnamefont {Drigo}}, \bibinfo {author}
  {\bibfnamefont {M.}~\bibnamefont {Bristow}}, \bibinfo {author} {\bibfnamefont
  {A.~J.}\ \bibnamefont {Schofield}}, \ and\ \bibinfo {author} {\bibfnamefont
  {A.~I.}\ \bibnamefont {Coldea}},\ }\href@noop {} {\enquote {\bibinfo {title}
  {Quenched nematic criticality separating two superconducting domes in an
  iron-based superconductor under pressure},}\ }\bibinfo {note} {To
  appear}\BibitemShut {NoStop}%
\bibitem [{\citenamefont {Hosoi}\ \emph {et~al.}(2016)\citenamefont {Hosoi},
  \citenamefont {Matsuura}, \citenamefont {Ishida}, \citenamefont {Wang},
  \citenamefont {Mizukami}, \citenamefont {Watashige}, \citenamefont
  {Kasahara}, \citenamefont {Matsuda},\ and\ \citenamefont
  {Shibauchi}}]{hosoi2016}%
  \BibitemOpen
  \bibfield  {author} {\bibinfo {author} {\bibfnamefont {S.}~\bibnamefont
  {Hosoi}}, \bibinfo {author} {\bibfnamefont {K.}~\bibnamefont {Matsuura}},
  \bibinfo {author} {\bibfnamefont {K.}~\bibnamefont {Ishida}}, \bibinfo
  {author} {\bibfnamefont {H.}~\bibnamefont {Wang}}, \bibinfo {author}
  {\bibfnamefont {Y.}~\bibnamefont {Mizukami}}, \bibinfo {author}
  {\bibfnamefont {T.}~\bibnamefont {Watashige}}, \bibinfo {author}
  {\bibfnamefont {S.}~\bibnamefont {Kasahara}}, \bibinfo {author}
  {\bibfnamefont {Y.}~\bibnamefont {Matsuda}}, \ and\ \bibinfo {author}
  {\bibfnamefont {T.}~\bibnamefont {Shibauchi}},\ }\href@noop {} {\bibfield
  {journal} {\bibinfo  {journal} {Proceedings of the National Academy of
  Sciences}\ }\textbf {\bibinfo {volume} {113}},\ \bibinfo {pages} {8139}
  (\bibinfo {year} {2016})}\BibitemShut {NoStop}%
\bibitem [{\citenamefont {Urata}\ \emph {et~al.}(2016)\citenamefont {Urata},
  \citenamefont {Tanabe}, \citenamefont {Huynh}, \citenamefont {Oguro},
  \citenamefont {Watanabe},\ and\ \citenamefont {Tanigaki}}]{Urata:2016aa}%
  \BibitemOpen
  \bibfield  {author} {\bibinfo {author} {\bibfnamefont {T.}~\bibnamefont
  {Urata}}, \bibinfo {author} {\bibfnamefont {Y.}~\bibnamefont {Tanabe}},
  \bibinfo {author} {\bibfnamefont {K.~K.}\ \bibnamefont {Huynh}}, \bibinfo
  {author} {\bibfnamefont {H.}~\bibnamefont {Oguro}}, \bibinfo {author}
  {\bibfnamefont {K.}~\bibnamefont {Watanabe}}, \ and\ \bibinfo {author}
  {\bibfnamefont {K.}~\bibnamefont {Tanigaki}},\ }\href
  {https://arxiv.org/pdf/1608.01044} {\bibfield  {journal} {\bibinfo  {journal}
  {arXiv:1608.01044}\ } (\bibinfo {year} {2016})}\BibitemShut {NoStop}%
\bibitem [{\citenamefont {Reiss}\ \emph {et~al.}(2017)\citenamefont {Reiss},
  \citenamefont {Watson}, \citenamefont {Kim}, \citenamefont {Haghighirad},
  \citenamefont {Woodruff}, \citenamefont {Bruma}, \citenamefont {Clarke},\
  and\ \citenamefont {Coldea}}]{reiss17}%
  \BibitemOpen
  \bibfield  {author} {\bibinfo {author} {\bibfnamefont {P.}~\bibnamefont
  {Reiss}}, \bibinfo {author} {\bibfnamefont {M.~D.}\ \bibnamefont {Watson}},
  \bibinfo {author} {\bibfnamefont {T.~K.}\ \bibnamefont {Kim}}, \bibinfo
  {author} {\bibfnamefont {A.~A.}\ \bibnamefont {Haghighirad}}, \bibinfo
  {author} {\bibfnamefont {D.~N.}\ \bibnamefont {Woodruff}}, \bibinfo {author}
  {\bibfnamefont {M.}~\bibnamefont {Bruma}}, \bibinfo {author} {\bibfnamefont
  {S.~J.}\ \bibnamefont {Clarke}}, \ and\ \bibinfo {author} {\bibfnamefont
  {A.~I.}\ \bibnamefont {Coldea}},\ }\href {\doibase
  10.1103/PhysRevB.96.121103} {\bibfield  {journal} {\bibinfo  {journal} {Phys.
  Rev. B}\ }\textbf {\bibinfo {volume} {96}},\ \bibinfo {pages} {121103}
  (\bibinfo {year} {2017})}\BibitemShut {NoStop}%
\bibitem [{\citenamefont {Oganesyan}\ \emph {et~al.}(2001)\citenamefont
  {Oganesyan}, \citenamefont {Kivelson},\ and\ \citenamefont
  {Fradkin}}]{oganesyan2001}%
  \BibitemOpen
  \bibfield  {author} {\bibinfo {author} {\bibfnamefont {V.}~\bibnamefont
  {Oganesyan}}, \bibinfo {author} {\bibfnamefont {S.~A.}\ \bibnamefont
  {Kivelson}}, \ and\ \bibinfo {author} {\bibfnamefont {E.}~\bibnamefont
  {Fradkin}},\ }\href@noop {} {\bibfield  {journal} {\bibinfo  {journal} {Phys.
  Rev. B}\ }\textbf {\bibinfo {volume} {64}},\ \bibinfo {pages} {195109}
  (\bibinfo {year} {2001})}\BibitemShut {NoStop}%
\bibitem [{\citenamefont {Metzner}\ \emph {et~al.}(2003)\citenamefont
  {Metzner}, \citenamefont {Rohe},\ and\ \citenamefont
  {Andergassen}}]{metzner03}%
  \BibitemOpen
  \bibfield  {author} {\bibinfo {author} {\bibfnamefont {W.}~\bibnamefont
  {Metzner}}, \bibinfo {author} {\bibfnamefont {D.}~\bibnamefont {Rohe}}, \
  and\ \bibinfo {author} {\bibfnamefont {S.}~\bibnamefont {Andergassen}},\
  }\href {\doibase 10.1103/PhysRevLett.91.066402} {\bibfield  {journal}
  {\bibinfo  {journal} {Phys. Rev. Lett.}\ }\textbf {\bibinfo {volume} {91}},\
  \bibinfo {pages} {066402} (\bibinfo {year} {2003})}\BibitemShut {NoStop}%
\bibitem [{\citenamefont {Lawler}\ \emph {et~al.}(2006)\citenamefont {Lawler},
  \citenamefont {Barci}, \citenamefont {Fern\'andez}, \citenamefont {Fradkin},\
  and\ \citenamefont {Oxman}}]{lawler06}%
  \BibitemOpen
  \bibfield  {author} {\bibinfo {author} {\bibfnamefont {M.~J.}\ \bibnamefont
  {Lawler}}, \bibinfo {author} {\bibfnamefont {D.~G.}\ \bibnamefont {Barci}},
  \bibinfo {author} {\bibfnamefont {V.}~\bibnamefont {Fern\'andez}}, \bibinfo
  {author} {\bibfnamefont {E.}~\bibnamefont {Fradkin}}, \ and\ \bibinfo
  {author} {\bibfnamefont {L.}~\bibnamefont {Oxman}},\ }\href {\doibase
  10.1103/PhysRevB.73.085101} {\bibfield  {journal} {\bibinfo  {journal} {Phys.
  Rev. B}\ }\textbf {\bibinfo {volume} {73}},\ \bibinfo {pages} {085101}
  (\bibinfo {year} {2006})}\BibitemShut {NoStop}%
\bibitem [{\citenamefont {Lee}(2009)}]{sslee09}%
  \BibitemOpen
  \bibfield  {author} {\bibinfo {author} {\bibfnamefont {S.-S.}\ \bibnamefont
  {Lee}},\ }\href {\doibase 10.1103/PhysRevB.80.165102} {\bibfield  {journal}
  {\bibinfo  {journal} {Phys. Rev. B}\ }\textbf {\bibinfo {volume} {80}},\
  \bibinfo {pages} {165102} (\bibinfo {year} {2009})}\BibitemShut {NoStop}%
\bibitem [{\citenamefont {Fradkin}\ \emph {et~al.}(2010)\citenamefont
  {Fradkin}, \citenamefont {Kivelson}, \citenamefont {Lawler}, \citenamefont
  {Eisenstein},\ and\ \citenamefont {Mackenzie}}]{fradkin10}%
  \BibitemOpen
  \bibfield  {author} {\bibinfo {author} {\bibfnamefont {E.}~\bibnamefont
  {Fradkin}}, \bibinfo {author} {\bibfnamefont {S.~A.}\ \bibnamefont
  {Kivelson}}, \bibinfo {author} {\bibfnamefont {M.~J.}\ \bibnamefont
  {Lawler}}, \bibinfo {author} {\bibfnamefont {J.~P.}\ \bibnamefont
  {Eisenstein}}, \ and\ \bibinfo {author} {\bibfnamefont {A.~P.}\ \bibnamefont
  {Mackenzie}},\ }\href@noop {} {\bibfield  {journal} {\bibinfo  {journal}
  {Annual Review of Condensed Matter Physics}\ }\textbf {\bibinfo {volume}
  {1}},\ \bibinfo {pages} {153} (\bibinfo {year} {2010})}\BibitemShut {NoStop}%
\bibitem [{\citenamefont {Metlitski}\ and\ \citenamefont
  {Sachdev}(2010)}]{metlitski10a}%
  \BibitemOpen
  \bibfield  {author} {\bibinfo {author} {\bibfnamefont {M.~A.}\ \bibnamefont
  {Metlitski}}\ and\ \bibinfo {author} {\bibfnamefont {S.}~\bibnamefont
  {Sachdev}},\ }\href {\doibase 10.1103/PhysRevB.82.075127} {\bibfield
  {journal} {\bibinfo  {journal} {Phys. Rev. B}\ }\textbf {\bibinfo {volume}
  {82}},\ \bibinfo {pages} {075127} (\bibinfo {year} {2010})}\BibitemShut
  {NoStop}%
\bibitem [{\citenamefont {Mross}\ \emph {et~al.}(2010)\citenamefont {Mross},
  \citenamefont {McGreevy}, \citenamefont {Liu},\ and\ \citenamefont
  {Senthil}}]{mross10}%
  \BibitemOpen
  \bibfield  {author} {\bibinfo {author} {\bibfnamefont {D.~F.}\ \bibnamefont
  {Mross}}, \bibinfo {author} {\bibfnamefont {J.}~\bibnamefont {McGreevy}},
  \bibinfo {author} {\bibfnamefont {H.}~\bibnamefont {Liu}}, \ and\ \bibinfo
  {author} {\bibfnamefont {T.}~\bibnamefont {Senthil}},\ }\href {\doibase
  10.1103/PhysRevB.82.045121} {\bibfield  {journal} {\bibinfo  {journal} {Phys.
  Rev. B}\ }\textbf {\bibinfo {volume} {82}},\ \bibinfo {pages} {045121}
  (\bibinfo {year} {2010})}\BibitemShut {NoStop}%
\bibitem [{\citenamefont {Fernandes}\ \emph {et~al.}(2014)\citenamefont
  {Fernandes}, \citenamefont {Chubukov},\ and\ \citenamefont
  {Schmalian}}]{fernandes14}%
  \BibitemOpen
  \bibfield  {author} {\bibinfo {author} {\bibfnamefont {R.~M.}\ \bibnamefont
  {Fernandes}}, \bibinfo {author} {\bibfnamefont {A.~V.}\ \bibnamefont
  {Chubukov}}, \ and\ \bibinfo {author} {\bibfnamefont {J.}~\bibnamefont
  {Schmalian}},\ }\href {\doibase 10.1038/nphys2877} {\bibfield  {journal}
  {\bibinfo  {journal} {Nature Physics}\ }\textbf {\bibinfo {volume} {10}},\
  \bibinfo {pages} {97} (\bibinfo {year} {2014})}\BibitemShut {NoStop}%
\bibitem [{\citenamefont {Holder}\ and\ \citenamefont
  {Metzner}(2015)}]{holder15}%
  \BibitemOpen
  \bibfield  {author} {\bibinfo {author} {\bibfnamefont {T.}~\bibnamefont
  {Holder}}\ and\ \bibinfo {author} {\bibfnamefont {W.}~\bibnamefont
  {Metzner}},\ }\href {\doibase 10.1103/PhysRevB.92.041112} {\bibfield
  {journal} {\bibinfo  {journal} {Phys. Rev. B}\ }\textbf {\bibinfo {volume}
  {92}},\ \bibinfo {pages} {041112} (\bibinfo {year} {2015})}\BibitemShut
  {NoStop}%
\bibitem [{\citenamefont {Karahasanovic}\ and\ \citenamefont
  {Schmalian}(2016)}]{Karahasanovic2016}%
  \BibitemOpen
  \bibfield  {author} {\bibinfo {author} {\bibfnamefont {U.}~\bibnamefont
  {Karahasanovic}}\ and\ \bibinfo {author} {\bibfnamefont {J.}~\bibnamefont
  {Schmalian}},\ }\href {\doibase 10.1103/PhysRevB.93.064520} {\bibfield
  {journal} {\bibinfo  {journal} {Phys. Rev. B}\ }\textbf {\bibinfo {volume}
  {93}},\ \bibinfo {pages} {064520} (\bibinfo {year} {2016})}\BibitemShut
  {NoStop}%
\bibitem [{\citenamefont {Paul}\ and\ \citenamefont {Garst}(2017)}]{Paul2017}%
  \BibitemOpen
  \bibfield  {author} {\bibinfo {author} {\bibfnamefont {I.}~\bibnamefont
  {Paul}}\ and\ \bibinfo {author} {\bibfnamefont {M.}~\bibnamefont {Garst}},\
  }\href {\doibase 10.1103/PhysRevLett.118.227601} {\bibfield  {journal}
  {\bibinfo  {journal} {Phys. Rev. Lett.}\ }\textbf {\bibinfo {volume} {118}},\
  \bibinfo {pages} {227601} (\bibinfo {year} {2017})}\BibitemShut {NoStop}%
\bibitem [{\citenamefont {Lederer}\ \emph {et~al.}(2015)\citenamefont
  {Lederer}, \citenamefont {Schattner}, \citenamefont {Berg},\ and\
  \citenamefont {Kivelson}}]{lederer15}%
  \BibitemOpen
  \bibfield  {author} {\bibinfo {author} {\bibfnamefont {S.}~\bibnamefont
  {Lederer}}, \bibinfo {author} {\bibfnamefont {Y.}~\bibnamefont {Schattner}},
  \bibinfo {author} {\bibfnamefont {E.}~\bibnamefont {Berg}}, \ and\ \bibinfo
  {author} {\bibfnamefont {S.~A.}\ \bibnamefont {Kivelson}},\ }\href {\doibase
  10.1103/PhysRevLett.114.097001} {\bibfield  {journal} {\bibinfo  {journal}
  {Phys. Rev. Lett.}\ }\textbf {\bibinfo {volume} {114}},\ \bibinfo {pages}
  {097001} (\bibinfo {year} {2015})}\BibitemShut {NoStop}%
\bibitem [{\citenamefont {Berg}\ \emph {et~al.}(2019)\citenamefont {Berg},
  \citenamefont {Lederer}, \citenamefont {Schattner},\ and\ \citenamefont
  {Trebst}}]{Berg:2018aa}%
  \BibitemOpen
  \bibfield  {author} {\bibinfo {author} {\bibfnamefont {E.}~\bibnamefont
  {Berg}}, \bibinfo {author} {\bibfnamefont {S.}~\bibnamefont {Lederer}},
  \bibinfo {author} {\bibfnamefont {Y.}~\bibnamefont {Schattner}}, \ and\
  \bibinfo {author} {\bibfnamefont {S.}~\bibnamefont {Trebst}},\ }\href@noop {}
  {\bibfield  {journal} {\bibinfo  {journal} {Annual Review of Condensed Matter
  Physics}\ }\textbf {\bibinfo {volume} {10}} (\bibinfo {year}
  {2019})}\BibitemShut {NoStop}%
\bibitem [{\citenamefont {Gallais}\ and\ \citenamefont
  {Paul}(2016)}]{gallais16}%
  \BibitemOpen
  \bibfield  {author} {\bibinfo {author} {\bibfnamefont {Y.}~\bibnamefont
  {Gallais}}\ and\ \bibinfo {author} {\bibfnamefont {I.}~\bibnamefont {Paul}},\
  }\href {\doibase https://doi.org/10.1016/j.crhy.2015.10.001} {\bibfield
  {journal} {\bibinfo  {journal} {Comptes Rendus Physique}\ }\textbf {\bibinfo
  {volume} {17}},\ \bibinfo {pages} {113 } (\bibinfo {year}
  {2016})}\BibitemShut {NoStop}%
\bibitem [{\citenamefont {Klein}\ \emph
  {et~al.}(2018{\natexlab{b}})\citenamefont {Klein}, \citenamefont {Lederer},
  \citenamefont {Chowdhury}, \citenamefont {Berg},\ and\ \citenamefont
  {Chubukov}}]{klein18b}%
  \BibitemOpen
  \bibfield  {author} {\bibinfo {author} {\bibfnamefont {A.}~\bibnamefont
  {Klein}}, \bibinfo {author} {\bibfnamefont {S.}~\bibnamefont {Lederer}},
  \bibinfo {author} {\bibfnamefont {D.}~\bibnamefont {Chowdhury}}, \bibinfo
  {author} {\bibfnamefont {E.}~\bibnamefont {Berg}}, \ and\ \bibinfo {author}
  {\bibfnamefont {A.}~\bibnamefont {Chubukov}},\ }\href {\doibase
  10.1103/PhysRevB.97.155115} {\bibfield  {journal} {\bibinfo  {journal} {Phys.
  Rev. B}\ }\textbf {\bibinfo {volume} {97}},\ \bibinfo {pages} {155115}
  (\bibinfo {year} {2018}{\natexlab{b}})}\BibitemShut {NoStop}%
\bibitem [{\citenamefont {Wang}\ and\ \citenamefont {Berg}(2019)}]{xw2019}%
  \BibitemOpen
  \bibfield  {author} {\bibinfo {author} {\bibfnamefont {X.}~\bibnamefont
  {Wang}}\ and\ \bibinfo {author} {\bibfnamefont {E.}~\bibnamefont {Berg}},\
  }\href {\doibase 10.1103/PhysRevB.99.235136} {\bibfield  {journal} {\bibinfo
  {journal} {Phys. Rev. B}\ }\textbf {\bibinfo {volume} {99}},\ \bibinfo
  {pages} {235136} (\bibinfo {year} {2019})}\BibitemShut {NoStop}%
\bibitem [{\citenamefont {Hartnoll}\ \emph {et~al.}(2018)\citenamefont
  {Hartnoll}, \citenamefont {Lucas},\ and\ \citenamefont {Sachdev}}]{HQM}%
  \BibitemOpen
  \bibfield  {author} {\bibinfo {author} {\bibfnamefont {S.}~\bibnamefont
  {Hartnoll}}, \bibinfo {author} {\bibfnamefont {A.}~\bibnamefont {Lucas}}, \
  and\ \bibinfo {author} {\bibfnamefont {S.}~\bibnamefont {Sachdev}},\
  }\href@noop {} {\emph {\bibinfo {title} {Holographic quantum matter}}}\
  (\bibinfo  {publisher} {The MIT Press},\ \bibinfo {year} {2018})\BibitemShut
  {NoStop}%
\bibitem [{\citenamefont {Hartnoll}\ \emph {et~al.}(2014)\citenamefont
  {Hartnoll}, \citenamefont {Mahajan}, \citenamefont {Punk},\ and\
  \citenamefont {Sachdev}}]{hartnoll14}%
  \BibitemOpen
  \bibfield  {author} {\bibinfo {author} {\bibfnamefont {S.~A.}\ \bibnamefont
  {Hartnoll}}, \bibinfo {author} {\bibfnamefont {R.}~\bibnamefont {Mahajan}},
  \bibinfo {author} {\bibfnamefont {M.}~\bibnamefont {Punk}}, \ and\ \bibinfo
  {author} {\bibfnamefont {S.}~\bibnamefont {Sachdev}},\ }\href
  {http://dx.doi.org/10.1103/PhysRevB.89.155130} {\bibfield  {journal}
  {\bibinfo  {journal} {Physical Review B}\ }\textbf {\bibinfo {volume} {89}}
  (\bibinfo {year} {2014})}\BibitemShut {NoStop}%
\bibitem [{\citenamefont {Schattner}\ \emph {et~al.}(2016)\citenamefont
  {Schattner}, \citenamefont {Lederer}, \citenamefont {Kivelson},\ and\
  \citenamefont {Berg}}]{schattner16}%
  \BibitemOpen
  \bibfield  {author} {\bibinfo {author} {\bibfnamefont {Y.}~\bibnamefont
  {Schattner}}, \bibinfo {author} {\bibfnamefont {S.}~\bibnamefont {Lederer}},
  \bibinfo {author} {\bibfnamefont {S.~A.}\ \bibnamefont {Kivelson}}, \ and\
  \bibinfo {author} {\bibfnamefont {E.}~\bibnamefont {Berg}},\ }\href {\doibase
  10.1103/PhysRevX.6.031028} {\bibfield  {journal} {\bibinfo  {journal} {Phys.
  Rev. X}\ }\textbf {\bibinfo {volume} {6}},\ \bibinfo {pages} {031028}
  (\bibinfo {year} {2016})}\BibitemShut {NoStop}%
\bibitem [{\citenamefont {Metlitski}\ \emph {et~al.}(2015)\citenamefont
  {Metlitski}, \citenamefont {Mross}, \citenamefont {Sachdev},\ and\
  \citenamefont {Senthil}}]{mross15}%
  \BibitemOpen
  \bibfield  {author} {\bibinfo {author} {\bibfnamefont {M.~A.}\ \bibnamefont
  {Metlitski}}, \bibinfo {author} {\bibfnamefont {D.~F.}\ \bibnamefont
  {Mross}}, \bibinfo {author} {\bibfnamefont {S.}~\bibnamefont {Sachdev}}, \
  and\ \bibinfo {author} {\bibfnamefont {T.}~\bibnamefont {Senthil}},\ }\href
  {\doibase 10.1103/PhysRevB.91.115111} {\bibfield  {journal} {\bibinfo
  {journal} {Phys. Rev. B}\ }\textbf {\bibinfo {volume} {91}},\ \bibinfo
  {pages} {115111} (\bibinfo {year} {2015})}\BibitemShut {NoStop}%
\bibitem [{\citenamefont {Forster}(2018)}]{forster}%
  \BibitemOpen
  \bibfield  {author} {\bibinfo {author} {\bibfnamefont {D.}~\bibnamefont
  {Forster}},\ }\href@noop {} {\emph {\bibinfo {title} {Hydrodynamic
  Fluctuations, Broken Symmetry, And Correlation Functions}}}\ (\bibinfo
  {publisher} {CRC Press; 1 edition},\ \bibinfo {year} {March 8,
  2018})\BibitemShut {NoStop}%
\bibitem [{\citenamefont {Maslov}\ \emph {et~al.}(2011)\citenamefont {Maslov},
  \citenamefont {Yudson},\ and\ \citenamefont {Chubukov}}]{maslov11}%
  \BibitemOpen
  \bibfield  {author} {\bibinfo {author} {\bibfnamefont {D.~L.}\ \bibnamefont
  {Maslov}}, \bibinfo {author} {\bibfnamefont {V.~I.}\ \bibnamefont {Yudson}},
  \ and\ \bibinfo {author} {\bibfnamefont {A.~V.}\ \bibnamefont {Chubukov}},\
  }\href {\doibase 10.1103/PhysRevLett.106.106403} {\bibfield  {journal}
  {\bibinfo  {journal} {Phys. Rev. Lett.}\ }\textbf {\bibinfo {volume} {106}},\
  \bibinfo {pages} {106403} (\bibinfo {year} {2011})}\BibitemShut {NoStop}%
\bibitem [{\citenamefont {Ledwith}\ \emph {et~al.}(2017)\citenamefont
  {Ledwith}, \citenamefont {Guo}, \citenamefont {Shytov},\ and\ \citenamefont
  {Levitov}}]{ledwidth17}%
  \BibitemOpen
  \bibfield  {author} {\bibinfo {author} {\bibfnamefont {P.~J.}\ \bibnamefont
  {Ledwith}}, \bibinfo {author} {\bibfnamefont {H.}~\bibnamefont {Guo}},
  \bibinfo {author} {\bibfnamefont {A.~V.}\ \bibnamefont {Shytov}}, \ and\
  \bibinfo {author} {\bibfnamefont {L.}~\bibnamefont {Levitov}},\ }\href
  {https://arxiv.org/pdf/1708.02376} {\bibfield  {journal} {\bibinfo  {journal}
  {arXiv:1708.02376}\ } (\bibinfo {year} {2017})}\BibitemShut {NoStop}%
\bibitem [{\citenamefont {de~Carvalho}\ and\ \citenamefont
  {Fernandes}(2019)}]{carvalho2019a}%
  \BibitemOpen
  \bibfield  {author} {\bibinfo {author} {\bibfnamefont {V.~S.}\ \bibnamefont
  {de~Carvalho}}\ and\ \bibinfo {author} {\bibfnamefont {R.~M.}\ \bibnamefont
  {Fernandes}},\ }\href {\doibase 10.1103/PhysRevB.100.115103} {\bibfield
  {journal} {\bibinfo  {journal} {Phys. Rev. B}\ }\textbf {\bibinfo {volume}
  {100}},\ \bibinfo {pages} {115103} (\bibinfo {year} {2019})}\BibitemShut
  {NoStop}%
\bibitem [{\citenamefont {Vieira}\ \emph {et~al.}(2020)\citenamefont {Vieira},
  \citenamefont {{de Carvalho}},\ and\ \citenamefont {Freire}}]{carvalho2019b}%
  \BibitemOpen
  \bibfield  {author} {\bibinfo {author} {\bibfnamefont {L.~E.}\ \bibnamefont
  {Vieira}}, \bibinfo {author} {\bibfnamefont {V.~S.}\ \bibnamefont {{de
  Carvalho}}}, \ and\ \bibinfo {author} {\bibfnamefont {H.}~\bibnamefont
  {Freire}},\ }\href {\doibase https://doi.org/10.1016/j.aop.2020.168230}
  {\bibfield  {journal} {\bibinfo  {journal} {Annals of Physics}\ }\textbf
  {\bibinfo {volume} {419}},\ \bibinfo {pages} {168230} (\bibinfo {year}
  {2020})}\BibitemShut {NoStop}%
\end{thebibliography}%
\end{document}